\documentclass{optica-article}

\journal{opticajournal} 

\articletype{Research Article}

\usepackage{lineno}
\usepackage{amsmath}
\usepackage{tabularx}

\begin{document}

\title{Subtleties of nanophotonic lithium niobate waveguides for on-chip evanescent wave sensing}

\author{
Nathan A. Harper,\authormark{1,$\dag$}
Emily Y. Hwang,\authormark{2,$\dag$}
Philip A. Kocheril,\authormark{1,$\dag$}
Tze King Lam,\authormark{3}
and Scott K. Cushing\authormark{1,*}
}

\address{
\authormark{1}Division of Chemistry and Chemical Engineering, California Institute of Technology, Pasadena, California 91125, USA\\
\authormark{2}Department of Applied Physics and Materials Science, California Institute of Technology, Pasadena, California, 91125, USA\\
\authormark{3}Deparment of Physics, St. Catharine’s College, University of Cambridge, Cambridge, CB2 1TN, UK\\
\authormark{$\dag$}These authors contributed equally to this work.
}

\email{\authormark{*}scushing@caltech.edu}


\begin{abstract*} 
Thin-film lithium niobate is a promising photonic platform for on-chip optical sensing because both nonlinear and linear components can be fabricated within one integrated device. To date, waveguided sample interactions for thin-film lithium niobate are not well explored. Compared to other integrated platforms, lithium niobate's high refractive index, birefringence, and angled sidewalls present unique design challenges for evanescent wave sensing. Here, we compare the performance of the quasi-transverse-electric (TE) and the quasi-transverse-magnetic (TM) mode for sensing on a thin-film lithium niobate rib waveguide with a 5~mM dye-doped polymer cladding pumped at 406~nm. We determine that both modes have propagation losses dominated by scatter, and that the absorption due to the sample only accounts for 3\% of the measured losses for both modes. The TM mode has better overlap with the sample than the TE mode, but the TM mode also has a stronger propagation loss due to sidewall and sample induced scattering (32.5~$\pm$~0.3~dB/cm) compared to the TE mode (23.0~$\pm$~0.2~dB/cm). The TE mode is, therefore, more appropriate for sensing. Our findings have important implications for on-chip lithium niobate-based sensor designs.
\end{abstract*}

\section{Introduction}

Thin-film lithium niobate (TFLN) is a promising candidate for on-chip sensing because its strong nonlinearities and low material losses allow for light generation, manipulation, and sample interaction within the same compact device\cite{boes_lithium_2023,zhu_integrated_2021}. Lithium niobate's strong quadratic nonlinearity, sub-$\mu$m modal confinement, and ability for quasi-phase matching lead to efficient frequency conversion through second harmonic generation\cite{jankowski_ultrabroadband_2020}, optical parametric amplification\cite{ledezma_intense_2022}, optical parametric oscillation \cite{lu_ultralow_threshold_2021, hwang_midIR_2023}, and spontaneous parametric downconversion\cite{xue_ultrabright_2021, Harper_2024}. TFLN has been used to generate light spanning from the UV-A\cite{hwang_tunable_2023} to the mid-IR\cite{mishra_mid-infrared_2021} in photonic circuits, with efficiencies unmatched in any other platform. Ultrafast laser pulses can also be generated on-chip \cite{zhang_broadband_2019,yu_integrated_2022, guo_ultrafast_2023} to produce white light with broad spectra from cascaded higher-order interactions\cite{wu_visible-to-ultraviolet_2024}. In addition to being an ideal on-chip light source, TFLN's nonlinearities are also useful for producing compact spectrometers\cite{pohl_spectrometer_2020, shams-ansari_dual-comb_2022} with integrated detectors\cite{sayem_SNSPD_2020}. 

While frequency conversion sources and modulation in TFLN are being widely explored, integrated sample interaction geometries for on-chip sensors are less studied. Since the evanescent field of light coupled into a waveguide extends only a few hundred nanometers past its surface, evanescent field sensing allows for highly specific analyte interrogation \cite{kozma2014integrated,benito2016fluorescence, kocheril2022amplification}. Evanescent wave sensors have been implemented in materials such as silicon, silicon nitride, and glass with multiple waveguide architectures, including fiber optic waveguides, planar waveguides, slot waveguides, rib waveguides, and strip waveguides\cite{wang2020optical,lavers2000planar,yuan2015mach,pi2023ultra}. For example, TE slot waveguides and TM strip waveguides have been proposed as optimal sensing geometries on silicon-on-insulator (SOI) based on the modal confinement factors and scattering losses of these waveguide configurations in the SOI platform\cite{kita_are_2018}. However, the material properties of lithium niobate present additional concerns for evanescent wave sensing designs. Lithium niobate's angled sidewall profile make slot waveguides difficult to fabricate, and its high refractive index and birefringence present unique design challenges to sample interaction geometries, warranting additional study. Evanescent wave sensing can be modeled with mode overlap calculations between the waveguide mode and the sample to investigate the efficiency of the sample excitation. However, practical devices need to account for additional sources of loss that are not well captured by modeling, such as the scattering from sub-wavelength imperfections in the waveguide profile, scatterers in the sample, and the refractive index contrast at the material interfaces.

Here, we quantify the interaction strength of the guided modes with a sample in a model TFLN waveguide fluorescence sensor. By measuring the fluorescence and scatter along the length of the device, we determine the amount of light absorbed by the sample compared to the amount of light scattered from the waveguide. The TFLN waveguide is an X-cut rib waveguide clad with a 60~nm thick 5~mM dye-doped polymer film and pumped with a 406~nm diode laser. Theoretical models predict that the fundamental TM mode of the rib waveguides have a nearly two-fold larger overlap with the sample than the quasi-transverse-electric (TE) mode. Experimentally, the TM mode is more efficiently absorbed by the sample than the TE mode (propagation losses of $0.95 \pm 0.01$ dB/cm and $0.67 \pm 0.01$ dB/cm, respectively). However, the propagation losses due to scatter are significantly higher for the TM mode compared to the TE mode ($31.5 \pm 0.3$ dB/cm and $22.4 \pm 0.2$ dB/cm, respectively). The TM mode thus loses its apparent advantage because of the realities of non-ideal waveguide sidewalls and sample nonuniformities. The TE mode within an X-cut TFLN film accesses the strongest optical and electro-optic nonlinearities, so the finding that the TE mode can be directly used for sensing is important for sensor compatibility with integrated frequency conversion sources and modulators. Our comparison of theory and experiment also provides important correction factors for modeling realistic waveguide properties in TFLN waveguide sensors.

\begin{figure}[hbtp!]
\centering\includegraphics[width=8.4cm]{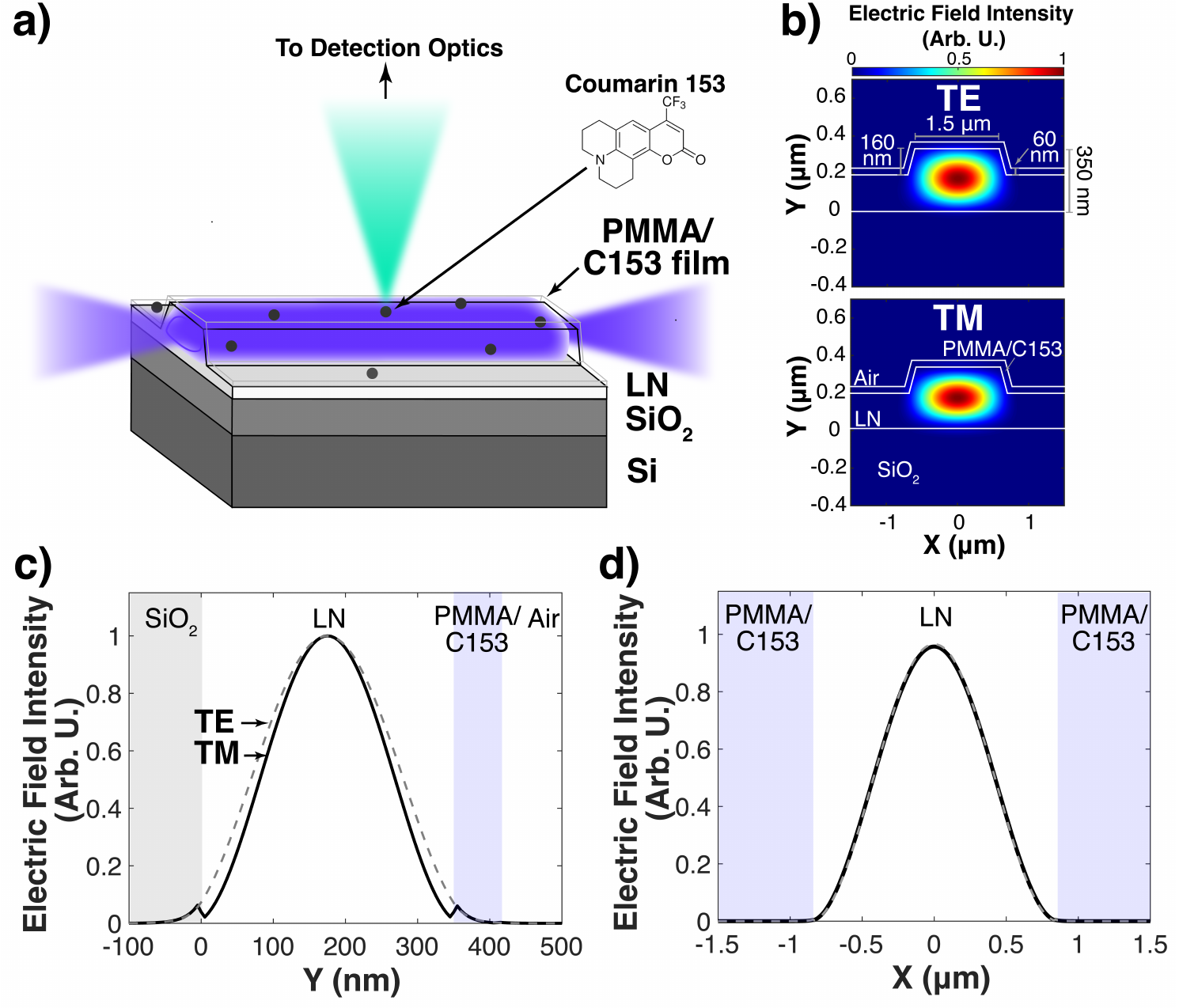}
\caption{\label{fig:theory} (a)~Schematic of the lithium niobate nanophotonic waveguides and PMMA/C153 film for evanescent wave sensing. (b)~Waveguide geometry and calculated electric field profiles profiles of the fundamental TE and TM modes using Lumerical. Lineouts of the simulated electric field in the (c)~$y$ direction in the center of the waveguide (at $x=0~\mu$m) and (d)~$x$ direction in the PMMA/C153 film (at $y=200$~nm).}
\end{figure}

\section{Device Design, Fabrication, and Characterization}
The waveguide coated with a dye-doped film was modeled in Lumerical MODE before fabrication (Figure~\ref{fig:theory}b-d). The confinement factors\cite{vlk_extraordinary_2021} and the propagation losses of the fundamental TE and TM waveguide modes are calculated in the presence of the the Coumarin-153 (C153) dye-doped polymethyl methacrylate (PMMA) film. For a 406~nm pump wavelength, waveguides with total lithium niobate thicknesses <400~nm have better sample-mode interaction overlaps (Supplemental Material Section S1) because a greater portion of the mode extends outside of the waveguide and into the dye-polymer layer as the waveguide cross sections decrease in area relative to the pump wavelength. For the waveguide geometry shown in Figure~\ref{fig:theory}b, the confinement factors of the fundamental TM and TE modes in the PMMA/C153 film are 0.013 and  0.0073, respectively. In a 60~nm homogeneous PMMA/C153 film (as experimentally measured through profilometry), the theoretical propagation loss due to the presence of the dye molecules is calculated as 7.2~dB/cm for the TM mode and 4.4~dB/cm for the TE mode. The difference in propagation loss between the modes arises from the difference in the electric field profile between the two fundamental waveguide modes, and is related to the aspect ratio of the waveguides relative to the polarization of the fundamental guided modes (Figure~\ref{fig:theory}c-d). 

The waveguides were then fabricated from an 8~mm by 12~mm die of a 5\% MgO-doped X-cut thin-film lithium niobate on insulator wafer (NanoLN), which consists of 350~nm of lithium niobate bonded to 2~$\mu$m of silicon dioxide on a 0.4~mm silicon substrate. Waveguides were patterned through an electron beam lithography exposure with hydrogen silesoxquiane resist (Applied Quantum Materials Inc.) followed by argon inductively coupled plasma reactive ion etching. The final device geometry consists of a lithium niobate rib waveguide with a $1.5~\mu$m top width and 160~nm etch depth (Figure~\ref{fig:theory}b). The chip facets were manually polished to increase coupling efficiency, resulting in a final waveguide length of approximately 5~mm. 

The PMMA/C153 films were prepared after the waveguide fabrication. To form the films, $1~\mu$L of a 10~mM stock solution of Coumarin 153 (Millipore Sigma 546186) in dimethyl sulfoxide (Millipore Sigma 276855) was added to $99~\mu$L of a PMMA (Millipore Sigma 182230) solution (1\% in toluene; Millipore Sigma 244511). $20~\mu$L of the PMMA/C153 solution was deposited on the TFLN chip by spin coating (MicroNano Tools BSC-100) at 300~rpm for 10~s followed by 2000~rpm for 60~s. The spin coating results in a film thickness of $\sim$60~nm, confirmed through profilometry on a separate device and in good agreement with literature \cite{rowinska2015fabrication}. Assuming minimal evaporation of the dimethyl sulfoxide, the C153 concentration is approximately 5~mM in the film. A confocal fluorescence micrograph verified that the film was relatively homogeneous across the sample, though strong variations in the fluorescence at the waveguide sidewalls and surface were observed (Supplemental Material Section S3).

\begin{figure}[htbp!]
\centering\includegraphics[width=8.4cm]{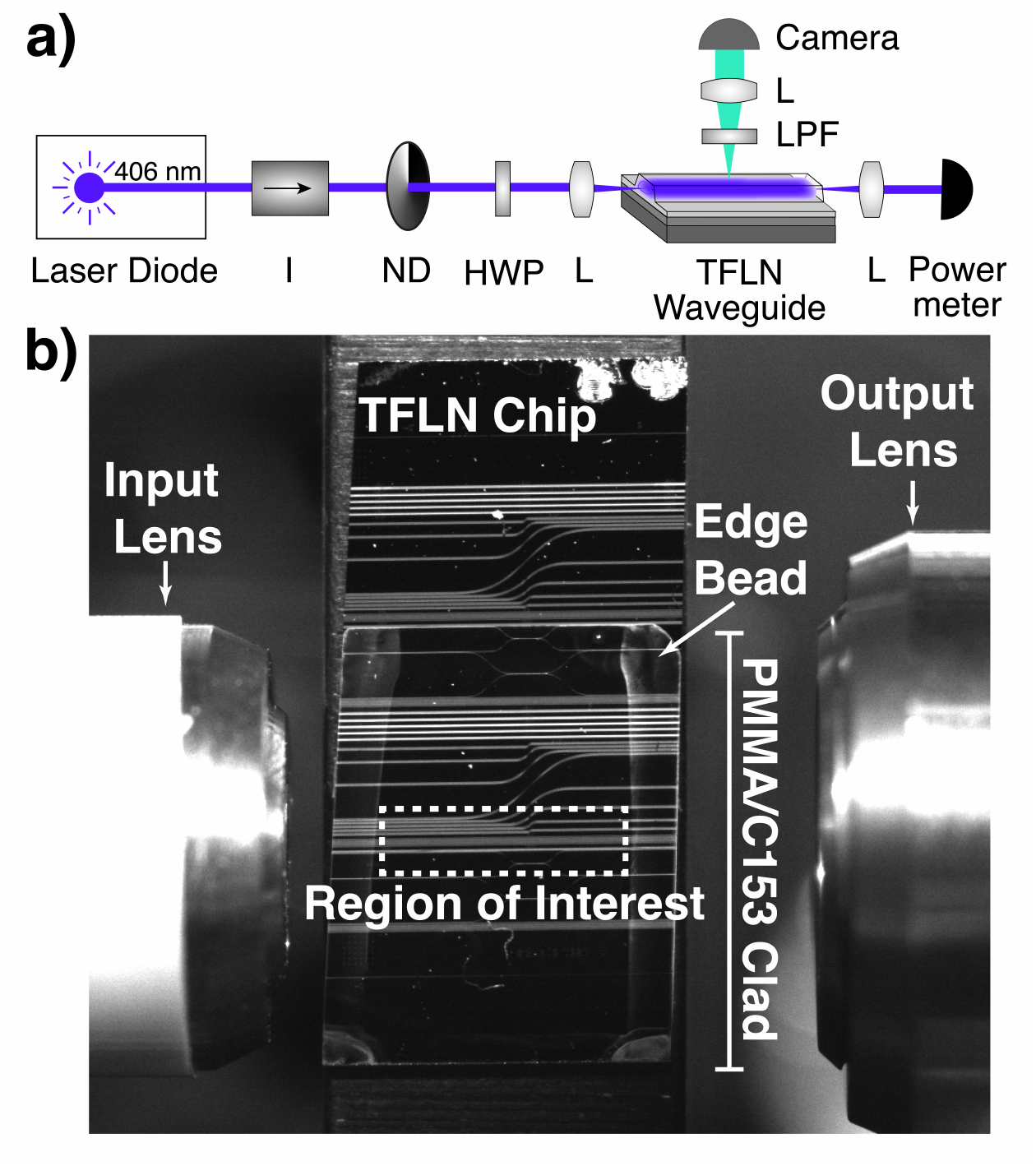}
\caption{\label{fig:setup} (a)~Optical setup for waveguide coupling and fluorescence detection. I, isolator; ND, variable neutral density filter; HWP, half-wave plate; L, lens; LPF, long-pass filter. (b)~Optical image of the lithium niobate chip and lens setup with the PMMA/C153 clad area, edge bead, and region of interest marked.}
\end{figure}

The fluorescence and scatter from the PMMA/C153 film on the waveguide are characterized as shown in Figure~\ref{fig:setup}a. In these experiments, the room-temperature waveguide is pumped with a free-running laser diode (Coherent OBIS LX 406 nm). An antireflection-coated aspheric lens (NA = 0.58, Thorlabs C140TMD-A) couples the free-space pump beam to the fundamental TE or TM mode of the waveguide depending on the polarization of the free-space beam, which is tuned with a half-wave plate. The total scatter from the top of the waveguides is collected using a camera lens (Edmund Optics 54691, $f=$75~mm, operated at $f/4$) and camera (Teledyne FLIR BFLY-U3-23S6M-C), with a collection efficiency of 0.025\%. To measure the fluorescence, a 500~nm long pass filter (Thorlabs FELH0500) is used with an optical density >5 at the pump wavelength while rejecting only 10\% of the generated fluorescence. A second aspheric lens collects the output of the waveguide into a power meter to monitor the stability of the coupling. Figure~\ref{fig:setup}b shows an image of the device and the coupling lenses used in this setup. All images were acquired for 5 minutes and were background corrected using images recorded with the laser shuttered. The images with the longpass filter were acquired for 1000~ms at 1~FPS (300~images total), while the images without the longpass filter were acquired for 100~ms at 10 FPS (3000 images total). To avoid photobleaching the dye, the alignment into the waveguide was performed at 5~nW of free-space power and image acquisition was performed with 100~nW of free-space power (Supplemental Material Section S5).

Due to the presence of an edge bead in the PMMA/C153 layer from the spin-coating process, a 4.3~mm section in the center of the waveguide with uniform film thickness (region of interest in Figure~\ref{fig:setup}b) is used for data analysis. The images are processed by summing the intensity of 24 vertical pixels at each horizontal pixel along the waveguide. The fluorescence counts are scaled up by 11\% to account for loss caused by the filter in the C153 emission spectrum passband (Supplemental Material Section S2). The counts from scatter are measured by subtracting the counts due to fluorescence from the counts recorded without the filter in place. A lithographically defined feature on the device is used to infer the measurement dimensions from the camera images.

\section{Results and Discussion} 
\begin{figure}[tb!]
\centering\includegraphics[width=6cm]{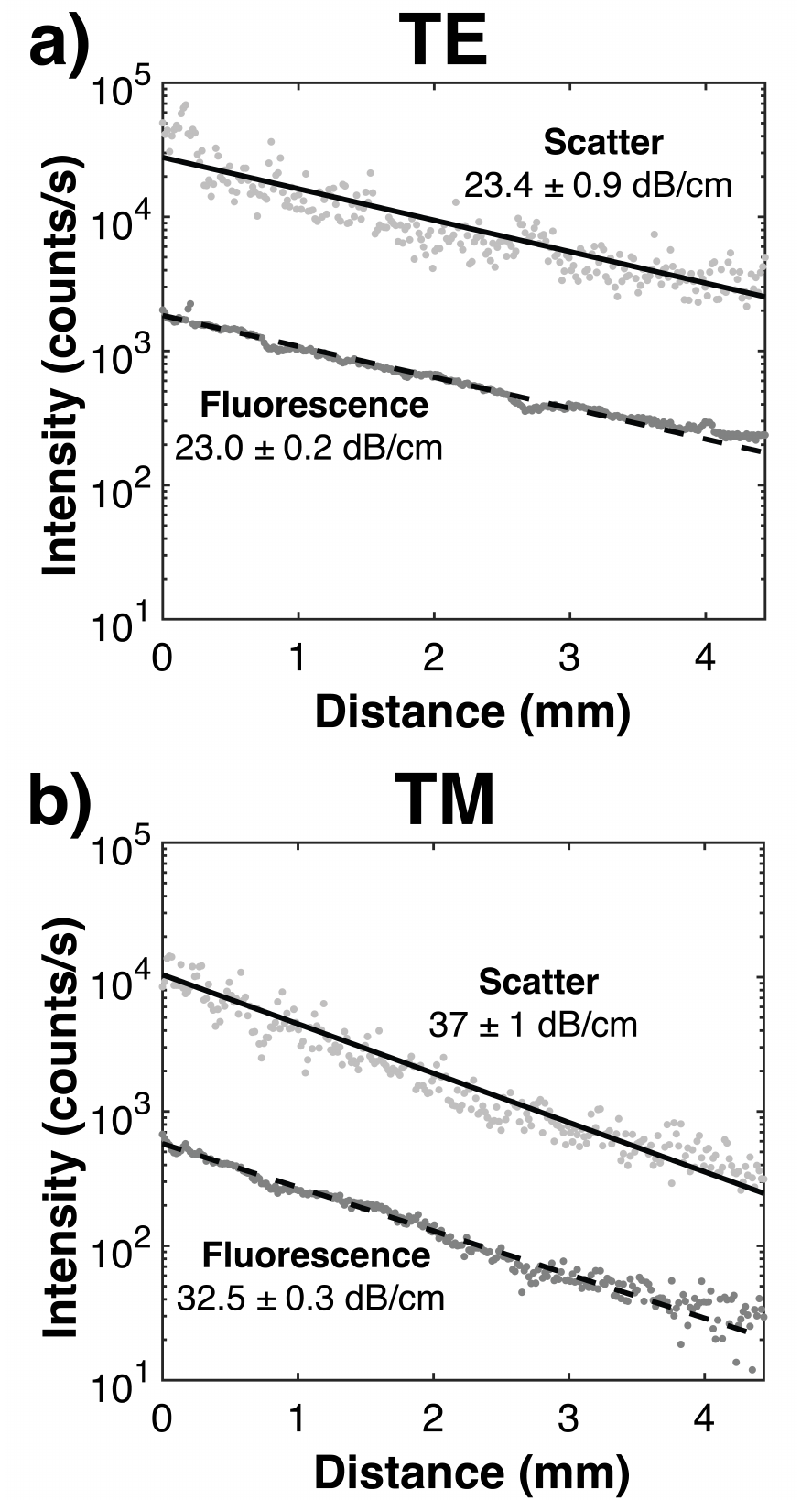}
\caption{\label{fig:loss} Measured fluorescence and scatter in the waveguide region of interest when coupled into the (a)~TE and (b)~TM modes.}
\end{figure}

Figures~\ref{fig:loss}a and \ref{fig:loss}b show the intensities measured across the images of the PMMA/C153-cladded waveguide when the TE and TM modes are excited. The traces are fit with a single exponential of the form $Ae^{-\alpha x}$, where $x$ is length, $\alpha$ is the propagation loss of the device with units of length$^{-1}$, and $A$ is a scaling term. The propagation loss of the TE mode is $23.0 \pm 0.2$ dB/cm, while the propagation loss of the TM mode is $32.5 \pm 0.3$ dB/cm, both inferred from the fluorescence images. A similar measurement using a bare waveguide determined that the propagation loss is $16.1 \pm 0.5$ dB/cm for the TE mode, and $18.1 \pm 0.7$ dB/cm for the TM mode, as summarized in Table~\ref{table:1}. The PMMA/C153 cladding therefore introduces additional loss to the waveguide.

Figures~\ref{fig:loss}a and \ref{fig:loss}b show that the intensity of the scattered signal is larger than the intensity of the fluorescence signal, which suggests that scattering is a more efficient loss mechanism than sample absorption in the device. The propagation loss measured from the fluorescence and the scatter shows good agreement for the TE mode ($23.0 \pm 0.2$ dB/cm and $23.4 \pm 0.9$ dB/cm, respectively) but show some disagreement for the TM mode ($32.5 \pm 0.3$ dB/cm and $37 \pm 1$ dB/cm, respectively), most likely due to the higher sensitivity of the TM mode to sidewall scattering and PMMA film variations. Since the fluorescence intensity shows less spatial variance than the scattering intensity (Figure~\ref{fig:loss} and Supplementary Material Figure~S4), the loss coefficient measured via fluorescence is chosen for all subsequent results.

Equation~\ref{derivative} is a modified form of Beer's Law, which models how the individual losses of sample absorption $\alpha_{\text{absorption}}$ and scattering $\alpha_{\text{scatter}}$ contribute to the total propagation loss $\alpha_{\text{total}}$ inferred from Figure~\ref{fig:loss}:

\begin{equation} \label{derivative}
    -\frac{1}{I(x)}\left(\frac{\textrm{d}I(x)}{\textrm{d}x}\right)  = \alpha_{\text{scatter}} + \alpha_{\text{absorption}} = \alpha_{\text{total}}
\end{equation}

\noindent Here, $I(x)$ is the intensity of the pump laser along the length of the waveguide. At each location along the waveguide, the number of photons lost to absorption $N_{\text{absorption}}(x)$ compared to the number of photons lost to scatter $N_{\text{scatter}}(x)$ is given by Equation~\ref{integration}:

\begin{equation} \label{integration}
    \frac{N_{\text{absorption}}(x)}{N_{\text{scatter}}(x)} = \frac{\alpha_{\text{absorption}}}{\alpha_{\text{scatter}}}
\end{equation}

\noindent The number of photons lost to absorption and scatter can be inferred from the data of Figure~\ref{fig:loss}. With the knowledge of the total propagation loss $\alpha_{\text{total}}$, the individual loss coefficients of absorption and scatter are then given by Equations~\ref{absorption} and~\ref{scatter}:

\begin{equation} \label{absorption}
    \alpha_{\text{absorption}} = \alpha_{\text{total}}\frac{N_{\text{absorption}}(x)}{N_{\text{absorption}}(x)+N_{\text{scatter}}(x)}
\end{equation}

\begin{equation} \label{scatter}
    \alpha_{\text{scatter}} = \alpha_{\text{total}}\frac{N_{\text{scatter}}(x)}{N_{\text{absorption}}(x)+N_{\text{scatter}}(x)}
\end{equation}

\noindent Since Equations~\ref{derivative}-\ref{scatter} hold over the entire length of the waveguide, the total number of photons lost through the region of interest $N_{\text{absorption}}$ and $N_{\text{scatter}}$ can be substituted for their spatially-dependent counterparts.

Over the 4.3~mm section of the waveguide used for loss measurements, the ratio of total counts due to scatter versus total counts due to fluorescence is $15.82 \pm 0.01$ and $15.73 \pm 0.04$ for the TE and TM modes, respectively. The raw integrated counts can be found in the Supplementary Material Section~S6. To relate the scatter-to-fluorescence ratio to the scatter-to-absorption ratio ($N_{\text{scatter}}/N_{\text{absorption}}$), the following factors are taken into account: 1) The collection efficiency of the system is the same for any photon radiated from the waveguide. 2) The camera quantum efficiency is 30\% at 406 nm and $\sim$70\% over the emission range of C153 (500-600 nm), as reported on the camera datasheet. 3) Fluorescence from the dye has a 90\% chance of transmitting through the filter based on the filter attenuation. 4) The quantum yield of Coumarin-153 in PMMA is 90\%, the same as its quantum yield in nonpolar solvents such as cyclohexane \cite{jones_solvent_1985}. 5) The PMMA film is thin enough that absorption of scatter and fluorescence reabsorption is negligible for both the TE and TM modes. With these assumptions, the probability that a photon lost to molecular absorption generates a count through fluorescence is roughly 2.1 times as high as the probability that a photon lost to scatter generates a count. The ratios of the scatter to molecular absorption for the TE and TM modes are therefore $33.22 \pm 0.02$ and $33.03 \pm 0.08$, respectively. The individual loss coefficients consistent with these ratios and the measured total propagation loss are displayed in Table~\ref{table:1}. The measured ratios confirm that scattering dominates the overall propagation loss in both the TE and TM modes. Compared to the TE mode, the TM mode is measured to experience higher loss, but not to a degree proportional to the fluorescence. Increased scattering in the TM mode is largely responsible for the increased total propagation loss, most likely due to inhomogeneities in the sample film. 

\begin{table}[b!]
\begin{center}
\begin{tabular}{ |c||c|c|c| } 
 \hline
Experiment & $\alpha_{\text{total}}$ (dB/cm) & $\alpha_{\text{scatter}}$ (dB/cm) & $\alpha_{\text{absorption}}$ (dB/cm) \\
 \hline \hline
PMMA/C153, TE & 23.0 $\pm$ 0.2 & 22.4 $\pm$ 0.2  & 0.67 $\pm$ 0.01  \\ \hline
PMMA/C153, TM & 32.5 $\pm$ 0.3  & 31.5 $\pm$ 0.3  & 0.95 $\pm$ 0.01  \\ \hline
Bare Waveguide, TE & 16.1 $\pm$ 0.5 & 16.1 $\pm$ 0.5 & 0 \\ \hline
Bare Waveguide, TM & 18.1 $\pm$ 0.7  & 18.1 $\pm$ 0.7 & 0 \\
 \hline
\end{tabular}
\caption{Summary of measured  and inferred propagation losses caused by scattering and C153 absorption.}
\label{table:1} 
\end{center}
\end{table}

The experiments confirm the theoretically predicted relative interaction strength of the TM mode compared to the TE mode with the sample (7.2~dB/cm and 4.4~dB/cm), but the measured interaction is much smaller (0.95~dB/cm and 0.67~dB/cm). The discrepancy comes from the difficulty in modeling the PMMA film parameters, especially the film profile in the vicinity of the waveguide. The PMMA/C153 layer was measured to be 60~nm on average using profilometry, but the film profile is not homogeneously 60~nm thick in the vicinity of the waveguide, as measured by confocal fluorescence microscopy (Supplementary Material Section~S3). The measured propagation loss could also be underestimated if the assumed dye quantum yield of 90\% is too high, which could be the case due to the PMMA microstructure around the dye, quenching of the dye (such as self-quenching by homo-F{\"o}rster resonance energy transfer), or degradation of the dye. Decreases in the dye cross-section from these effects by up to 50\% has been observed in the Coumarin family\cite{donovalova_PMMA_coumarin_2012} when dissolved in PMMA. If the dye quantum yield and film non-uniformity are included in the simulations, the absorption losses can become as low as 0.80~dB/cm for TM and 0.55~dB/cm for TE, close to the experimentally measured values in Table~\ref{table:1}. 

Based on our findings, future work improving TFLN-based evanescent field sensors should focus on decreasing the waveguide scattering loss imparted by the fabrication process and realistic sample interaction layers. Propagation losses of 0.002~dB/cm have been demonstrated in TFLN waveguides with methods such as post-fabrication annealing\cite{shams-ansari_reduced_2022}, ion beam milling\cite{siew_ultra-low_2018}, and redeposition-free etching\cite{kaufmann_redeposition-free_2023}. Compared to the losses in Table~\ref{table:1}, such improvements in the scattering loss will increase the fluorescence-to-scatter ratio from the 3\% demonstrated here to upwards of 30,000\%. A liquid sample with microfluidics will also eliminate much of the uncertainties associated with the polymer film used here. 

\section{Conclusion}
TFLN rib waveguides are analyzed for their light-matter interaction strength. The thin-film thickness is found to be the primary variable for increasing the sample interaction. While the fundamental TM mode is predicted to exhibit a two-fold stronger sample interaction, increased scattering losses from the waveguide for the TM mode outweigh this factor. We demonstrate the importance of quantifying different loss mechanisms on the waveguide when measuring a LN sensor's efficacy, which we accomplish by comparing the spatially-resolved intensities of scatter and fluorescence without requiring any knowledge of the actual on-chip pump intensities. The suitability of the TE versus TM mode for sensing has important implications. The main advantage of using TFLN is the integration of a sensor with up- or down-stream nonlinear frequency conversion, which utilizes the TE mode in X-cut TFLN to access the highest quadratic nonlinearity. Therefore, a sensor utilizing the TE mode in a rib waveguide is the most compatible with up- or downstream nonlinear frequency conversion. Although a polarization mode converter could be designed to convert between the TE and TM modes, any additional integrated components will likely incur more loss than is gained in interaction strength. Further improvements in the sensor performance, regardless of TE or TM mode, can be achieved with improved fabrication to decrease the scattering loss. Our work, therefore, represents a first step toward fully integrated TFLN sensors.

\begin{backmatter}
\bmsection{Funding} U.S. Department of Energy (DE-SC0022089)

\bmsection{Acknowledgments}
The authors gratefully acknowledge the critical support and infrastructure provided for this work by The Kavli Nanoscience Institute (KNI) and the Beckman Biological Imaging Facility at Caltech. This work was additionally supported by the KNI-Wheatley Scholar in Nanoscience and the Rothenberg Innovation Initiative. The authors thank Ryoto Sekine and Alireza Marandi for providing the TFLN dies and Giada Spigolon for her helpful discussions in fluorescence imaging. N.A.H. was supported by the Department of Defense (DoD) through the National Defense Science and Engineering Graduate (NDSEG) Fellowship Program. E.Y.H. and P.A.K. were supported by the National Science Foundation Graduate Research Fellowship Program under Grant no. DGE‐1745301. P.A.K. is grateful for financial support from a Hertz Fellowship. Any opinion, findings, and conclusions or recommendations expressed in this material are those of the authors(s) and do not necessarily reflect the views of the National Science Foundation. 

\bmsection{Disclosures}
The authors declare no conflicts of interest.

\bmsection{Data availability} Data underlying the results presented in this paper are not publicly available at this time but may be obtained from the authors upon reasonable request.

\bmsection{Supplemental Document} See the Supplemental Material for supporting content. 

\end{backmatter}

\bibliography{wgsensing}

\end{document}


\maketitle

\section{Electric Field Simulations}\label{section:theory}
\begin{figure}[b!]
\centering
\includegraphics[width=12cm]{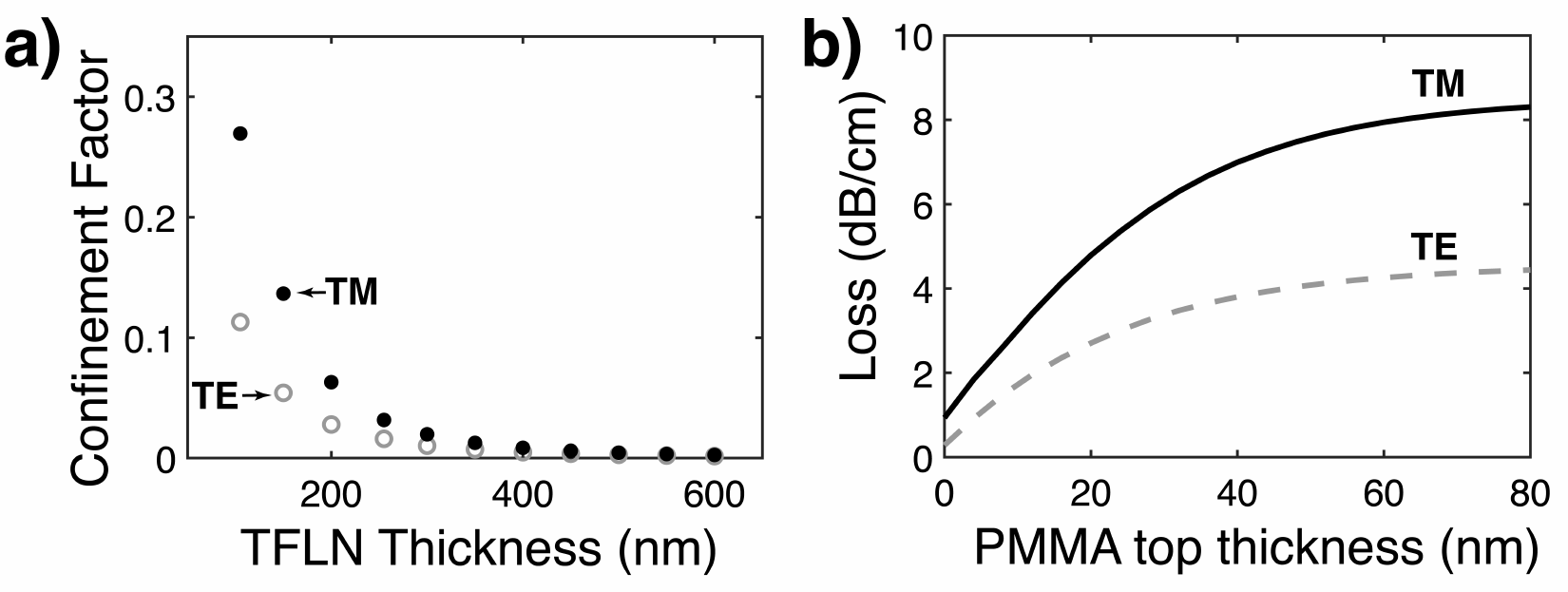}
\caption{a)~Confinement factor $\Gamma$ against TFLN thickness with a fixed waveguide aspect ratio and conformal 60~nm PMMA/C153 layer. b)~Theoretically predicted loss of the TE and TM modes with varying PMMA thickness on the top of the waveguide with the waveguide profile in Figure 1b of the main text.}
\label{fig:thickness}
\end{figure}

All simulations were performed in Lumerical MODE. The confinement factor provides a measure of the modal confinement in the PMMA/C153 sample layer and can be expressed as\cite{vlk_extraordinary_2021}: 
\begin{equation}
\Gamma = \frac{n_{\text{g}}}{\text{Re}\{n_{\text{clad}}\}}  \frac{\iint_\text{clad} \varepsilon |\textbf{E}|^{2} \,dx\,dy} {\iint_{-\infty} ^{+\infty} \varepsilon |\textbf{E}|^{2} \,dx\,dy}
\end{equation}
The confinement factor varies with the waveguide geometry, particularly the thin film thickness (Figure~\ref{fig:thickness}a).

To simulate the absorption of Coumarin-153, the dye loss can be incorporated into the simulated material parameters, where the refractive index of the PMMA/C153 is represented as $n' = n+i\kappa$. The real component of the refractive index is the index of bulk PMMA at the excitation wavelength, or $n = 1.5$. The imaginary index $\kappa$ incorporates the properties and concentration of Coumarin-153: 
\begin{equation}
\kappa = \frac{\alpha \lambda_{0}}{4\pi} = 2.303 \frac{\varepsilon c \lambda_{0}}{4\pi} = 0.000743
\end{equation}
where the cross section $\varepsilon$ is 20,000 cm$^{-1}$M$^{-1}$ for Coumarin-153 in toluene, the predicted dye concentration $c$ is 5~mM based on the solution concentrations, the excitation wavelength $\lambda_{0}$ is 406~nm, and the factor of 2.303 comes from the change of base. Incorporating the dye loss $\kappa$ in the simulation material parameters, the theoretical propagation loss due to the presence of the dye molecules is 4.4~dB/cm for the TE mode and 7.2~dB/cm for the TM mode with a 60~nm conformal PMMA/C153 film thickness. As discussed in the main text, these predicted losses do not match with the experimentally measured losses of 0.67~dB/cm for TE and 0.95~dB/cm for TM. The discrepancy in the predicted and measured losses likely arises from the unknown PMMA/C153 film quality, particularly the film profile near the waveguides. Consistent with this uncertainty, confocal fluorescence microscopy indicates that the PMMA/C153 layer is not uniform over the waveguide profile (Section~\ref{section:fluorescence}), particularly the top surface of the waveguides. Varying the PMMA/C153 thickness on top of the waveguides but maintaining 60~nm thickness elsewhere (Figure~\ref{fig:thickness}b), the predicted losses decrease to 1.1~dB/cm for TE and 1.6~dB/cm for TM, much closer to the measured values. The losses can be further decreased by incorporating a lower C153 quantum yield, which could be the case due to the PMMA microstructure around the dye, quenching of the dye (such as self-quenching by homo-F{\"o}rster resonance energy transfer), or degradation of the dye. Decreasing the quantum yield to 50\% produces theoretical losses of 0.80~dB/cm for TM and 0.55~dB/cm for TE, much closer to the experimentally measured losses of 0.95~dB/cm for TM and 0.67~dB/cm for TE. The remainder of the discrepancy can be attributed to uncertainties in the dye cross section, uncertainties in the dye concentration, and potential dye aggregation.

\section{Dye Properties}
\begin{figure}[hbp!]
\centering
\includegraphics[width=10cm]{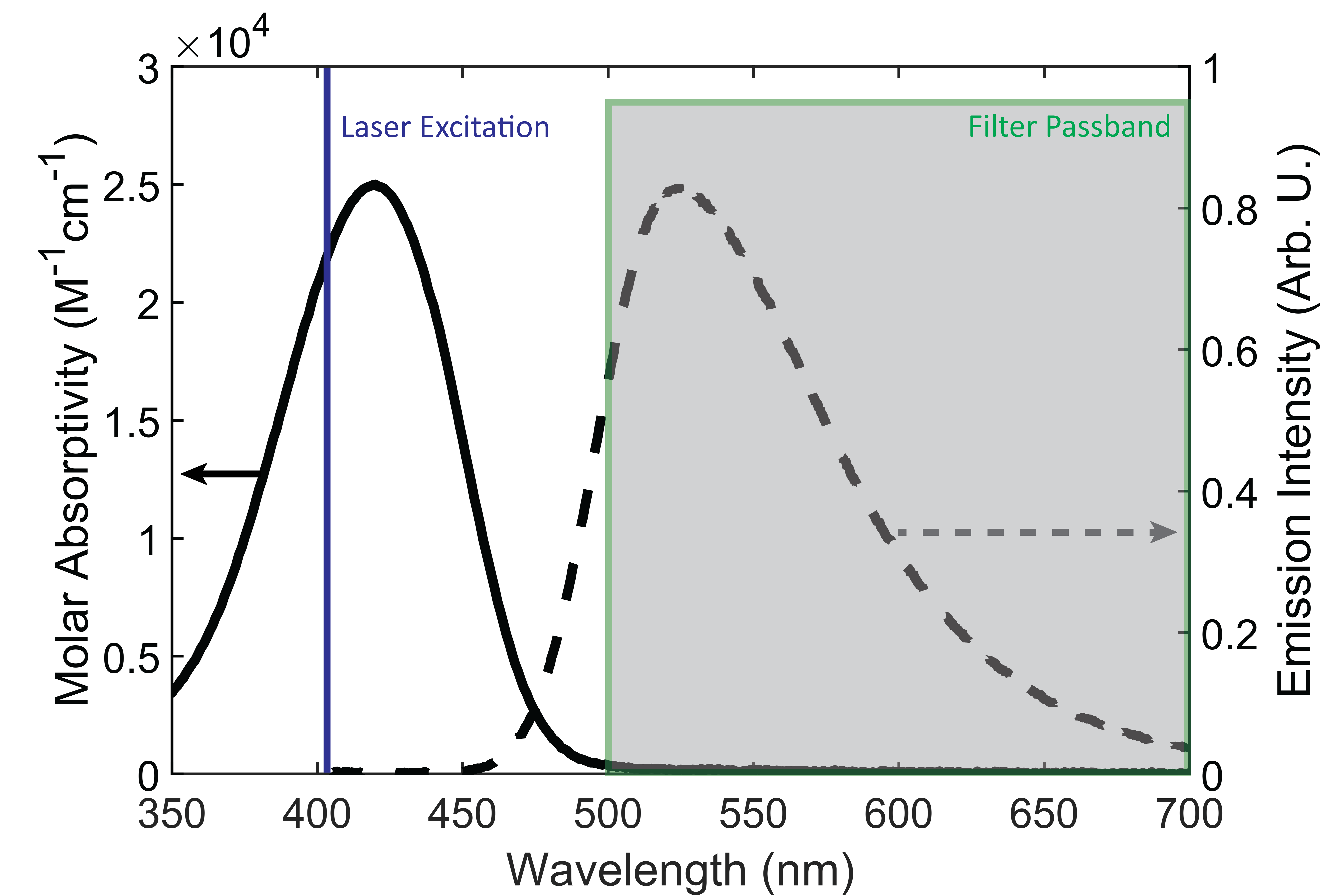}
\caption{Absorption and emission spectra of Coumarin-153 in DMSO with the laser excitation and filter passband wavelengths marked.}
\label{fig:spectrum}
\end{figure}

UV-vis absorption spectra were recorded with a Varian Cary 500 spectrophotometer (Agilent). Fluorescence emission spectra were recorded with an RF-6000 fluorimeter (Shimadzu). Spectra were acquired with a 10~$\mu$M stock solution of C153 in dimethyl sulfoxide (DMSO), and are shown in Figure~\ref{fig:spectrum}.

To estimate the percentage of the fluorescence that is able to transmit through the 500~nm longpass filter $T_{\text{fluorescence}}$, the following integrals over the emission intensity $I(\lambda)$ are numerically evaluated. The limits to these integrals are determined by the wavelength range of the instrument and the filter cutoff.

\begin{equation} \label{fluorescence_percent}
    T_{\text{fluorescence}} = \frac{\int_{\text{500 nm}}^{\text{928 nm}} I(\lambda) d\lambda}{\int_{\text{405 nm}}^{\text{928 nm}} I(\lambda) d\lambda} = 0.90
\end{equation}

\section{Fluorescence Microscopy}\label{section:fluorescence}
\begin{figure}[htbp!]
\centering
\includegraphics[width=13.2cm]{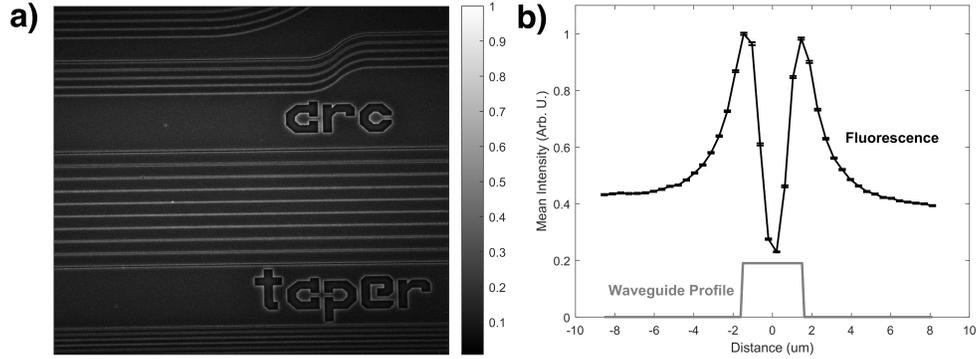}
\caption{a)~Fluorescence microscope image of the PMMA/C153 film on the waveguides. b)~Lineout of the fluorescence intensity across a waveguide.}
\label{fig:microscope_image}
\end{figure}

Fluorescence images of the device were recorded with a commercial scanning confocal microscope (Zeiss LSM 880). The sample was excited with a 405~nm laser diode in an epifluorescence scheme with a 20x dry objective. An image of the PMMA/C153 cladded device is shown in Figure~\ref{fig:microscope_image}a. The top-most straight waveguide is of the same dimensions as the waveguide investigated in the main text, though it is not the same waveguide due to concerns over photobleaching. A lineout of the intensity collected across the waveguide is shown in Figure~\ref{fig:microscope_image}b. Each data point is the mean count value at each vertical pixel acquired by integrating the image in the horizontal direction, and the error bars represent one standard error of the mean, calculated from the variance in the counts at each vertical pixel. From Figure~\ref{fig:microscope_image}b, it appears that the thickness of the PMMA film is not constant across the waveguide profile due to the strong variance in fluorescence as a function of position. Notably, the fluorescence is strongest near the etched areas of the waveguide, and lowest at the top surface of the waveguide. While the fluorescence profile could be explained by variations in the PMMA/C153 thickness, it could also be the result of differences in excitation and collection efficiencies from thin-film interference effects. We therefore suggest that the PMMA thickness might be spatially varying, but do not attempt to make any quantitative claims.

\section{Raw data figures with error bars}

\begin{figure}[H]
\centering
\includegraphics[width=13.3cm]{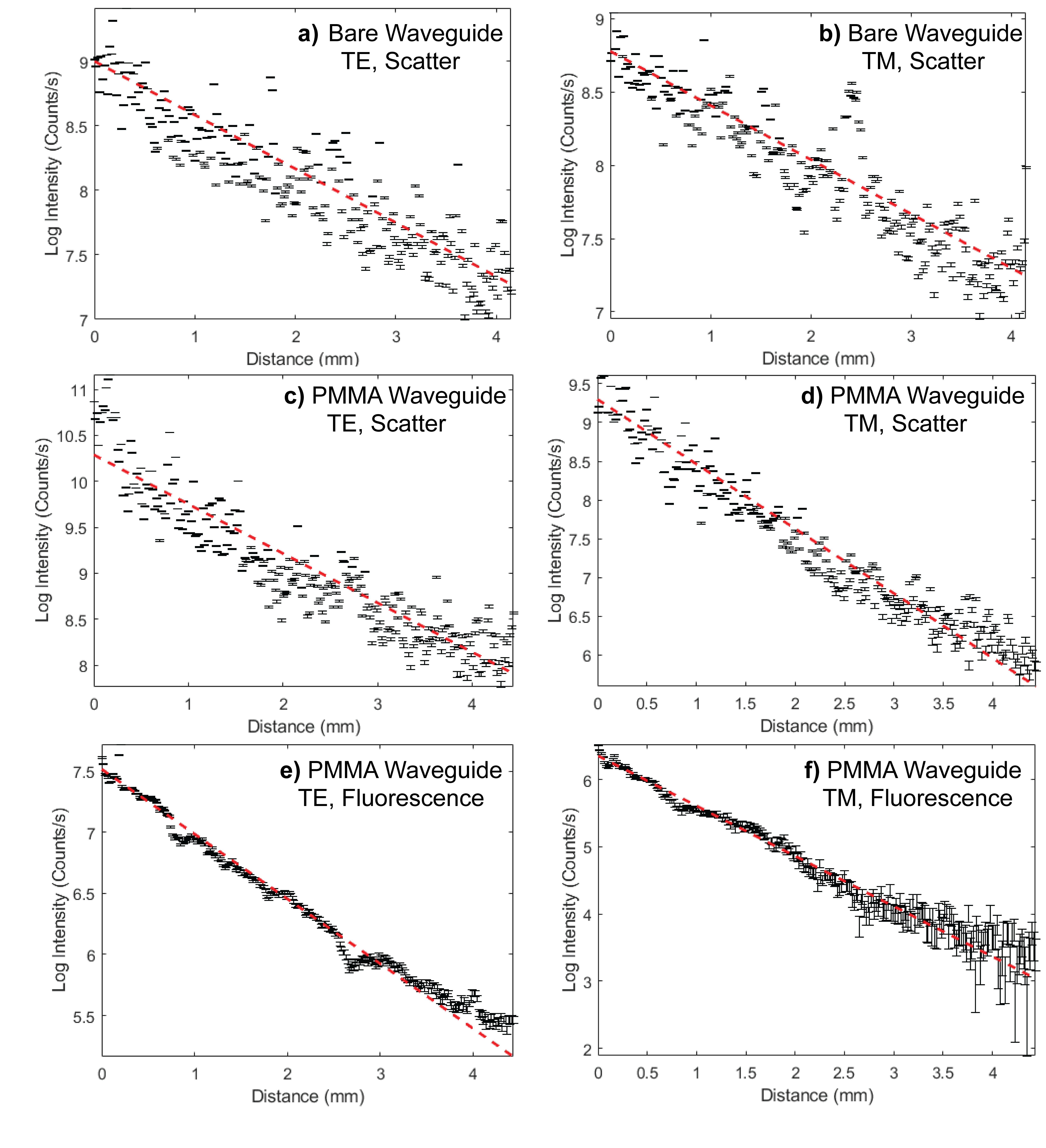}
\caption{Raw intensities for a-b) pump scatter from the bare waveguide, c-d) pump scatter from waveguide with PMMA cladding, e-f) fluorescence from waveguide with PMMA cladding. Each data point represents a 17~$\mu$m section of waveguide. The red dashed line is a linear fit to the logarithmic data.}
\label{fig:rawdata}
\end{figure}

The raw data from the imaging experiments are displayed in Figure~\ref{fig:rawdata}. Uncertainties are given as one standard error of the mean. The uncertainties from each figure are directly comparable as each experiment has the same total integration time of 5 minutes, and all images are acquired with the same digitizer gain (30~dB). The fluorescence datasets are acquired at 1 FPS with an integration time of 1 s, while the scatter datasets are acquired at 10 FPS with a 100 ms integration time to avoid saturating the camera. The uncertainty of the fluorescence data is larger than the scatter data due to the low light intensity, and is especially apparent at the end of the waveguide where the pump is most attenuated.

As discussed in the main text, the propagation loss measured from the fluorescence and the scatter shows good agreement for the TE mode ($23.0 \pm 0.2$ dB/cm and $23.4 \pm 0.9$ dB/cm, respectively) but show some disagreement for the TM mode ($32.5 \pm 0.3$ dB/cm and $37 \pm 1$ dB/cm, respectively), most likely due to the higher sensitivity of the TM mode to sidewall scattering and PMMA film variations. Potential reasons for the disagreement in the loss measurements of the TM mode are underestimations in the fit error or the presence of higher-order modes in the waveguide. The spatial variance in the scattering data of Figure~\ref{fig:rawdata} and Figure~3 of the main text is larger than that of the fluorescence, most likely due to uneven distribution of scattering sites or speckle interference from the narrow laser bandwidth. 

\section{Photobleaching analysis}

\begin{figure}[H]
\centering
\includegraphics[width=13.3cm]{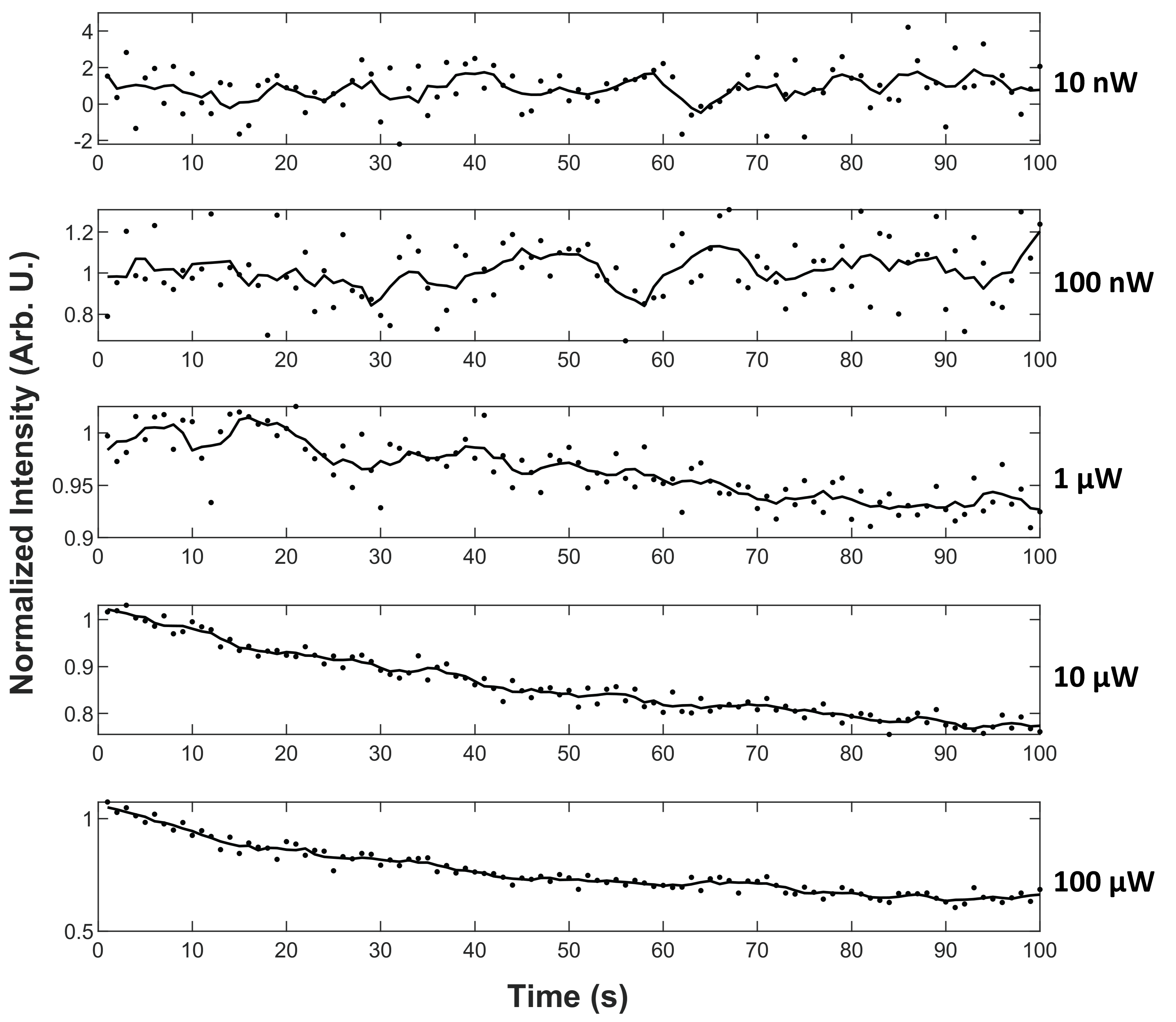}
\caption{Integrated fluorescence intensity over time for increasing values of free-space power. Data points show the raw intensity at each time point, and lines are a 5~s moving average. Intensities have been normalized relative to the mean intensity of the first 10 data points. }
\label{fig:bleach}
\end{figure}

The incident free space power into a waveguide was increased to determine the appropriate measurement settings with minimal photobleaching. Note that a different waveguide with identical dimensions to the one presented in the main text was tested here to avoid photobleaching the dye in the main device of interest. After coupling to the TM mode of the waveguide using 5~nW of incident power, the laser was set to the desired power, and fluorescence data were recorded for 100~s, followed by a 100~s background with the laser off. This process was repeated a total of five times, with the power level increasing by a factor of 10 for each measurement. A time series of the total amount of fluorescence collected over a 4~mm section of the waveguide are displayed in Figure~\ref{fig:bleach}. While it is difficult to determine if any photobleaching occurs at 10~nW or 100~nW due to the signal-to-noise of the measurement, photobleaching is apparent at 1~$\mu$W, and increases in strength at 10~$\mu$W and 100~$\mu$W. A power of 100~nW was chosen for the experiments in this work to maximize signal while keeping photobleaching to a relatively low level. 

\section{Raw data of aggregate counts}

\begin{table}[htpb!]
\begin{center}
\begin{tabular}{ |c||c| } 
 \hline
Experiment & Total Count Rate in ROI (kcps) \\
 \hline \hline
Bare Waveguide, TE, Scatter & 799.8 $\pm$ 0.3  \\ \hline
Bare Waveguide, TM, Scatter & 890.5 $\pm$ 0.3  \\ \hline
PMMA/C153, TE, Scatter & 3091 $\pm$ 1  \\ \hline
PMMA/C153, TM, Scatter & 735.9 $\pm$ 0.3  \\ \hline
PMMA/C153, TE, Fluorescence & 185.3 $\pm$ 0.1  \\ \hline
PMMA/C153, TM, Fluorescence & 43.7 $\pm$ 0.1  \\
 \hline
\end{tabular}
\caption{Summary of the total signal recorded for each experiment, integrated over a 4.3~mm section of the waveguide.}
\label{table:1} 
\end{center}
\end{table}

The total count rate over then entire device is summarized in Table~\ref{table:1}. Uncertainties are given as standard errors of the mean. These total counts are proportional to the amount of pump power in the ROI, in addition to the scattering and fluorescence loss rates. Since the pump power in the ROI region is dependent on the chip coupling efficiency and any losses in the edge beam region of the device, these numbers cannot be directly compared to each other. 

\bibliography{wgsensingsupplemental}